\documentclass[aps,prl,twocolumn,groupedaddress,floatfix,showpacs,showkeys,amsmath,amssymb]{revtex4-1}
\usepackage{graphicx}
\usepackage{bm}
\begin{document}
\title{Magnetism and metal-insulator transition in oxygen deficient SrTiO$_3$}

\author{Alejandro Lopez-Bezanilla$^{1}$}
\email[]{alejandrolb@gmail.com}
\author{P. Ganesh$^{2}$}
\author{Peter B. Littlewood$^{1}$ $^{3}$}
\affiliation{$^{1}$Argonne National Laboratory, 9700 S. Cass Avenue, Lemont, Illinois, 60439, United States}
\affiliation{$^{2}$Center for Nanophase Materials Science, Oak Ridge National Laboratory, One Bethel Valley Road, Tennessee, 37831, United States}
\affiliation{$^{3}$James Franck Institute, University of Chicago, Chicago, Illinois 60637, United States}

\begin{abstract}
First-principles calculations to study the electronic and magnetic properties of bulk, oxygen-deficient SrTiO$_3$ (STO) under different doping conditions and densities have been conducted.
The appearance of magnetism in oxygen-deficient STO is not determined solely by the presence of a single oxygen vacancy but by the density of free carriers and the relative proximity of the vacant sites.
We find that while an isolated vacancy behaves as a non-magnetic double donor, manipulation of the doping conditions allows the stability of a single donor state, with emergent local moments coupled ferromagnetically by carriers in the conduction band. Strong local lattice distortions enhance the binding of this state. The energy of the in-gap local moment can be further tuned by orthorhombic strain.
Consequently we find that the free-carrier density and strain are fundamental components to obtaining trapped spin-polarized electrons in oxygen-deficient STO, which may have important implications in the design of optical devices.

\end{abstract}

\maketitle	
Bulk SrTiO$_3$ (STO) at stoichiometry is a semiconductor with a substantial band gap and an empty d-shell. It has long been known that doping, particularly via Oxygen vacancies (Ov), leads to metallic behavior\cite{PhysRevLett.88.075508} and even superconductivity in the bulk\cite{PhysRev.163.380}. More recently, at controlled interfaces between insulating STO and LaAlO$_3$, a two-dimensional electron gas can be created \cite{ISI:000188470500037}\cite{ISI:000254144700022}. Also, confined in a thick layer at an open vacuum-cleaved surface of free-standing STO experimental evidence of charge carriers has been reported \cite{ISI:000286143400033}.
Quite surprising in view of the nearly empty d-shell has been the observation of robust ferromagnetism, coexisting with superconductivity. This has been seen in interfacial structures\cite{ISI:000247648000011,ISI:000306099900049}, but also in bulk and thin film , both in systems that are doped with other transition metals, but also those that are not. \cite{Middey,PhysRevLett.96.027207,Xu,PhysRevLett.88.075508, Crandles}
Recent advances in the synthesis of oxygen-deficient structures have allowed for the detection, measurement and control of optically induced long-lived magnetic moments in STO, further implicating Ov\cite{Crooker}. In consequence, it is important to understand the potential roles of Ov both in doping, and in the independent generation of local moments associated with the Ov.

In this study we perform first-principles calculations to show that the magnetic properties of single Ov in SrTiO$_3$ are richer than previously reported, exhibiting a large tunability with respect to the defect concentration and external doping rate. By modifying carrier-doping concentrations typical localized moments of high-density defects may vanish as the doping rate per unit of volume decreases, suggesting that external doping may restore the magnetism of the monovacancies and tune the magnetic moments associated with diluted Ov.
The occurrence of magnetic localized states in Ov in STO is intimately connected to the overlap of their electronic wavefunctions. Altering this overlap by the application of strain or modifying the density of free carriers may have profound effects on both the magnetism and the metallic/semiconducting character of the defected material. First-principles studies of magnetism for different unit cell parameters and doping rates in cubic oxygen-deficient STO provide a unique opportunity to tune, hence help decipher, complex exchange interaction between electron-donor defects in transition metal oxides of current technological and fundamental interest. The application of a wide range of tensile strains shifts the energy position of the Ov impurity bands and modulates the optical properties of the defected material.
Total density of states (DoS) and real-space projections of the net charge density will illustrate the tuning capability of oxygen-deficient STO with external physical stimuli.

The density functional theory based calculations were conducted using the projector augmented-wave method \cite{PhysRevB.50.17953} and the PBE-GGA exchange-correlation functional \cite{PhysRevLett.77.3865}. To improve the description of the electrons occupying the d-orbitals of the Ti atoms at the vacant site, a Hubbard-U correction (GGA+U) as implemented in the VASP code \cite{PhysRevB.48.13115,PhysRevB.54.11169,PhysRevB.59.1758} was included. The rotationally invariant method by Dudarev et al. \cite{PhysRevB.57.1505} with an effective U$_{eff}$=U-J=4.0 eV was applied to capture the strong correlations. The electronic wavefunctions were described using a plane-wave basis set with an energy cutoff of 400 eV. Atomic positions were fully relaxed in $\Gamma$-centered 3$\times$3$\times$3 supercells for k-point sampling until residual forces were lower than 0.05 eV/\AA. The number of k-points was decreased proportionally as the number of cells increased in the unit cell. Extra electrons were introduced or removed and compensated with an equally uniform background charge of opposite sign.

\begin{figure}[htp]
 \centering
 \includegraphics[width=0.45 \textwidth]{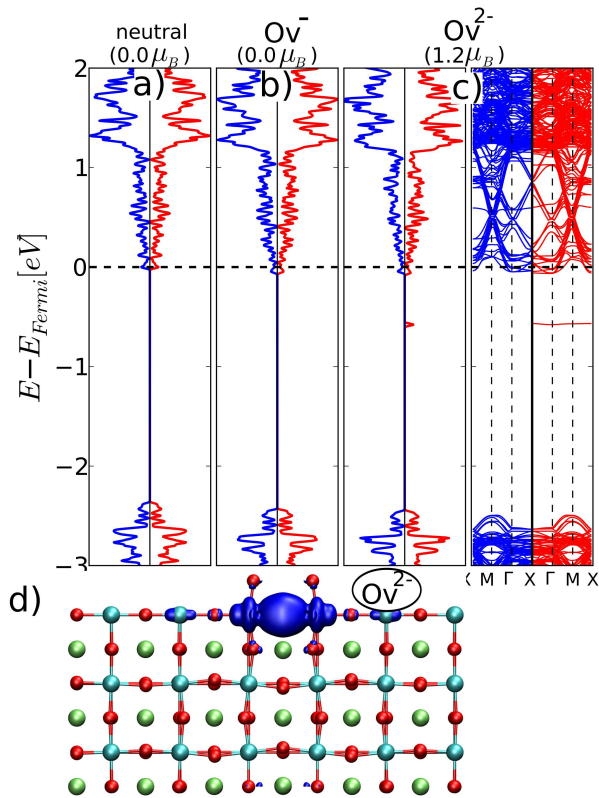}
 \caption{In a 3$\times$3$\times$6 supercell, we plot the density of states of a neutral (a) and singly- (b) and doubly-charged (c) vacancy. The doubly charged vacancy produces a localised moment (d) together with a strong lattice distortion.
 The excess of spin-up with respect to spin-down charge density in a  supercell is plotted in d) for an isosurface value of 0.01 e$^-$/\AA$^3$. }
 \label{fig1}
\end{figure}

In the cubic phase, STO is a non-magnetic wide-bandgap oxide material possessing the common features of perovskites, namely vertex-sharing metal-centered octahedra, square pyramids, and square planes. An oxygen vacancy is one of the intrinsic defects that may change STO's structural, magnetic and electronic properties as a result of the double-donor doping character of the defect. Indeed, an Ov is known to introduce electron carriers and cause an insulator-to-metal transition, as well as distortions in the atomic network yielding octahedra tilts \cite{ISI:000312985700008}. The result of removing an oxygen atom is the enlargement of the distance between the two Ti atoms to which it was bonded, and an overall reduction of the crystal symmetry from cubic to the C$_{4v}$ space group. The ground state of a single Ov in a 3$\times$3$\times$3 STO supercell is obtained upon expansion of 1\% of the cubic lattice parameter. As already  described, \cite{PhysRevLett.98.115503,PhysRevLett.111.217601,ISI:000312985700008,Terakura}, each Ti atom at the vacancy site has a dangling bond which further combine into molecular-like bonding and antibonding orbitals. These $\sigma$-orbitals are the result of the local hybridization of the Ti 3d$_{z^2-r^2}$ and 4p$_z$ orbitals,

In Figure \ref{fig1}a the DoS of a single Ov in a 3$\times$3$\times$6 supercell is plotted, where the longest dimension is in the direction of the Ti-Ov-Ti bond. Such a large unit cell turns out to be necessary to reproduce the physics of an isolated Ov. The neutral vacancy is seen to be a double donor, as conventionally assumed \cite{PhysRevLett.111.217601}, and the system exhibit a metalic behavior. We now consider the effect of adding extra electrons to the system (mimicking electrostatic doping). Adding a single electron (Figure \ref{fig1}b) simply shifts the Fermi level and leaves the system spin symmetric and paramagnetic. Adding a second electron (Figure \ref{fig1}c) produces a sudden restructuring of the density of states and a localised state with a full magnetic moment appears in the gap, and the ground state has a substantial ferromagnetic moment. Observing the spin density (Figure \ref{fig1}d), we note the strongly localized nature of this state at the vacancy position, together with substantial polaronic distortions. Adding a single electron to the Ov$^-$ configuration without allowing relaxation does {\em not} bind a localized state, namely the lattice relaxations are critical in the description of the magnetic configuration.
The electrostatic repulsion between this local charge density and the anionic near-neighbor O atom couples the localized state to a very localized atomic distortion, which is noticeable by 10$^\circ$ increase of the O-Ti-O angle around the Ov. This magnetic polaronic distortion is responsible for trapping the localized state by lowering its energy.  The strong polaronic coupling explains that ionizing oxygen-defective STO first removes electrons from the conduction band (CB) before the polaronic state is destroyed. Also to be noted is that the orbital configuration cannot be projected onto Ti d-states alone since the localised state lives in the vacancy itself. The metallic character of the oxygen-deficient STO so far described becomes semiconducting if the monovacancy is further isolated. Indeed, STO with a single Ov in a 5$\times$5$\times$5 supercell exhibits a band gap and a complete absence of magnetic state. Janotti et al.\cite{Janotti14} have recently pointed out the near-stability of polarons in the vicinity of an Ov.

\begin{figure}[htp]
 \centering
 \includegraphics[width=0.45 \textwidth]{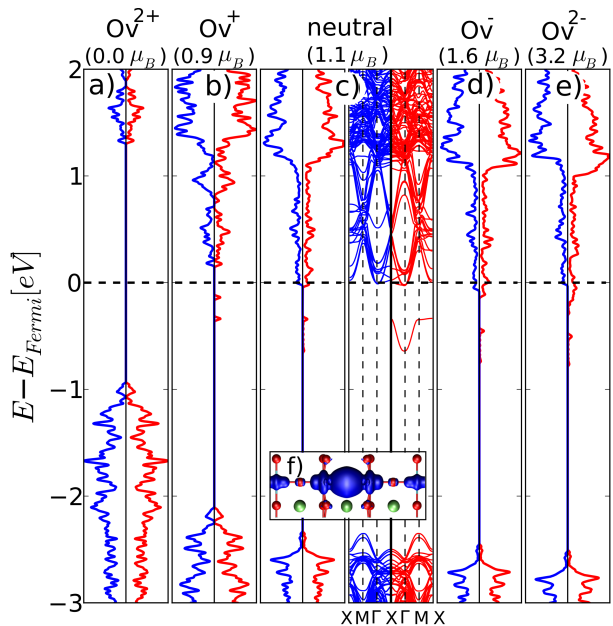}
 \caption{Total density of states of oxygen vacancies (Ov) in STO with increasing doping rates ranging from hole to electron doping. In a), two-fold hole-doped oxygen-defected STO recovers the wide band gap of bulk STO according with double-donor character of the single Ov. In b) two Ov in a double-size supercell two-fold hole-doped are in ferromagnetic coupling.  When doped with two extra e$^-$ the vacancy nearly reaches a half-metallic state. The excess of charge density for the neutral vacancy is plotted in f) for an isosurface value of 0.01 e$^-$/\AA$^3$ }
 \label{fig2}
\end{figure}

In Figure \ref{fig2}, we study a nearly identical system with a 3$\times$3$\times$3 STO supercell (i.e. twice the density and the doping rate of the above neutral Ov). Here the {\em neutral} vacancy yields a single fully occupied spin-polarized bonding in-gap state at $\sim$0.5 eV below the CB and a net spin-resolved charge density located in the surrounding of the vacant site (Figure\ref{fig2}f). The oxygen-deficient STO exhibits a magnetic moment of $\sim$1.1 $\mu_B$. This is in agreement with previous studies \cite{PhysRevLett.98.115503,PhysRevLett.111.217601,ISI:000312985700008,Terakura}, where the interpretation was given that correlation effects driven by the strength of the on-site Coulomb potential promote a second electron to the CB, confering the oxygen-deficient STO the metallic character observed in Figure \ref{fig2}c. Notice however the substantial dispersion of the in-gap state in Figure \ref{fig2}c which provides an explanation of the apparent discrepancy with the results of an isolated vacancy in Figure\ref{fig1}. This in-gap state of Figure \ref{fig2}exhibits a narrow energy spread in the k-space line joining the $M$ and $X$ points, related to a reduced overlap of the wavefunction between neighboring cells. On the contrary, a more dispersive character for the line joining the $M$ and $X$ points through the $\Gamma$ point indicates that the impurity band is extended in the Ti-Ov-Ti bond direction more than in the perpendicular directions, allowing the wavefunction to enhance the overlap with the adjacent vacancy states and develop a dispersive band.

Hole doping upon removal of one electron from the defected system leads to the suppression of the itinerant electron and the confinement of the remaining electron in the hydrogenic orbital, rendering the system a magnetic insulator, as shown in  Figure \ref{fig2}b. If two neighboring vacancies were allowed to interact in a larger two-fold hole-doped 3$\times$3$\times$6 supercell with the same doping rate imposing a total magnetic moment 	of zero, the moments would interact ferromagnetically forming a shallow in-gap state while the extra electron would go into the conduction band rendering the system half-metalic. The latter configuration is $\sim$100 meV per unit cell higher in energy than the former.
Further removal of an additional electron leads to two-fold hole doping and the oxygen-deficient STO to recover the wide band gap of the pristine bulk (see Figure \ref{fig2}a). On the contrary, electron doping enhances the metallic character as the additional electrons populate the half-filled d-bands. For high doping rates the unbalanced occupancy of the Ti d-orbitals leads the system towards a half-metallic state.

To recapitulate: An {\em isolated} Ov is a double donor, as has been traditionally assumed. However, when added electrons are located in the CB and the Ov wavefunctions overlap a magnetic configuration is stabilised, with each Ov a {\em single} donor (and therefore a local moment) coupled by spin-polarised itinerant carriers by double exchange, rendering the system ferromagnetic. A similar ferromagnetic state can also be obtained in an undoped but higher-density system of Ov, where the overlap between the vacancies stabilises the magnetism.

These results are in excellent agreement with the reported experimental evidence \cite{PhysRevLett.88.075508} of metallic conductivity in STO single crystals. This metallicity was removed after a treatment of reduction to induce a self-healing process and a decreasing of the density of initially introduced oxygen-vacancy defects and charge carriers. The consistency of our results with previous experimental observations allows us to conclude that the magnetic and electronic properties of high-dense arrays of Ov ranging from a total absence to highly localized magnetic moments are subjected to overlapping degree conditions, density of defects and doping rate.

\begin{figure*}[htp]
 \centering
 \includegraphics[width=0.95 \textwidth]{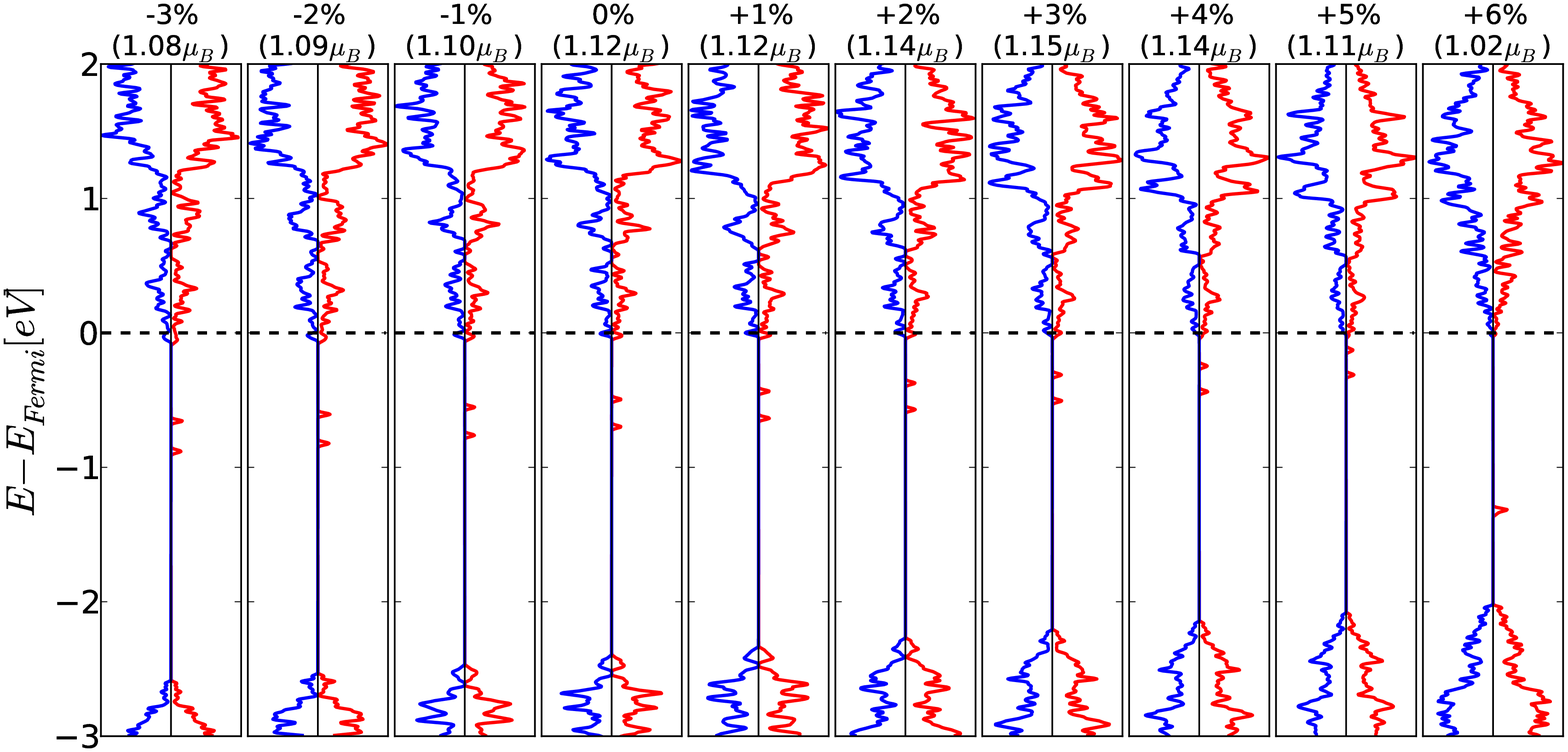}
 \caption{Evolution of the electronic states of oxygen-deficient STO under biaxial strain.}
 \label{fig4}
\end{figure*}

Many of the outstanding properties of transition metal oxides stem from the interactions between lattice, charge and spin degrees of freedom. Modifications on the lattice parameter as a result of tensile or compresive strain can be used to tune the electrical and magnetic properties of these materials. Typical strains within the range from -4\% to +10\% have been applied on epitaxially growth thin films and remarkably different properties were obtained. For instance, pristine STO remains paraelectric at very low temperatures and the application of stress results in ferroelectricity \cite{PhysRevB.13.271}. Epitaxial strain was used to perform a large increase of the STO transition temperature and produce room-temperature ferroelectricity \cite{ISI:000223233600035}, demonstrating that engineering strain provides an effective way to tune STO intrinsic properties and achieve novel functionalities. We next analyze the effect of strain on the electronic and magnetic properties of high-density oxygen-deficient STO.

Biaxial strain was modelled by means of successive modifications of two of the lattice parameter of a 3$\times$3$\times$3 supercell with a single Ov. This supercell size allows us to observe the evolution of the electronic and magnetic properties of the localized vacant state when tuning from compresive -3\% to tensile +6\% biaxial strain. We noticed that an expansion of 1\% of the defectless bulk STO lattice constant is required upon formation of the hybrid orbitals for stabilizing the formation energy of the vacant site. The magnetic moment is barely modified throughout the expansion until the transition to a new type of in-gap localized state at +6\%.
As the oxygen-deficient supercell varies from compressive to tensile biaxial strain at fixed composition and charge carrier stoichiometry, the energy of the in-gap state progressively evolves from $\sim$ -0.7 eV up to energy values close to the CB electrons. Therefore, the trapped electron evolves from a deep in-gap to a shallow state.
According to the general theory of impurity bands in n-type oxides \cite{ISI:000226749400024}, the former state fulfills the electronic structure diagram required by 3d-transition metal oxide based materials to exhibit low Curie temperature, whereas the proximity of the impurity band 3d-orbitals to the Fermi level of the former facilitates the high Curie temperature. A band gap reduction of $\sim$ 0.5 eV is also observed.

Light emission in irradiation-induced metallic oxygen-deficient STO \cite{ISI:000232931000008} exhibits an extraordinary tunability depending on the recombination between excited electrons or holes with conducting or defect-level electrons. The possibility to strain-engineering the oxygen vacancy state may lead to important applications on oxide-based electronic and optic devices that depend on the doped charge carrier concentrations and trapped electron energy level position with respect to the CB to modulate the luminescence induced by photo-excited carriers.

To sum up,
our findings explains that the formation of oxygen-vacancies along high density extended defected areas of STO may lead to metallic states. Vacancy states are magnetic for a sufficiently high density of free carriers achieved either by high density clustering of double-donor oxygen vacancies or appropriate amount of external doping.
The dynamics of the Ov in-gap state in response to external stimuli has been captured through DFT+U modelling showing that it provides an effective route to control and manipulate the magnetism and  electronic states at small scales.

We acknowledge the computing resources provided on Blues high-performance computing cluster operated by the Laboratory Computing Resource Center at Argonne National Laboratory. Work at Argonne is supported by DOE-BES under Contract No. DE-AC02-06CH11357. PG was sponsored by the laboratory Directed Research and Development Program of Oak Ridge National Laboratory, managed by UT-Battelle,LLC,for the US Department of Energy. Discussions with Andrew Millis and Scott Crooker are gratefully acknowledged

\end{document}